\documentclass[superscriptaddress, preprint, showpacs]{revtex4-1}
\usepackage{geometry}                % See geometry.pdf to learn the layout options. There are lots.
\usepackage[parfill]{parskip}    % Activate to begin paragraphs with an empty line rather than an indent
\usepackage{graphicx}
\usepackage{amssymb}
\usepackage{amsmath}
\usepackage{epstopdf}
\usepackage{subfigure}
\usepackage{color}
\newcommand{\pdag}{{\phantom{\dagger}}}
\newcommand{\ham}[1]{\mathcal{H}_{\text{#1}}}
\newcommand{\braket}[2]{\left\langle #1\,\right|\left. #2\right\rangle}
\newcommand{\bra}[1]{\left\langle #1\,\right|}
\newcommand{\ket}[1]{\left|\, #1\right\rangle}

\begin{document}

\title{Phase space analysis of interacting fermions}
\author{M. Ossadnik}
\affiliation{Institute for Theoretical Physics, ETH H\"onggerberg, CH-8093 Z\"urich, Switzerland}
\date{\today}                                           % Activate to display a given date or no date

\begin{abstract}
We propose the use of an orthogonal wave packet basis to analyze the low-energy physics of interacting electron systems with short range order. We give an introduction to wave packets and the related phase space representation of fermion systems, and show that they lend themselves to an efficient description of short range order. We illustrate the approach within an RG calculation for the one-dimensional Hubbard chain. 
\end{abstract}


\maketitle

%\tableofcontents

\section{Introduction}

Renormalization group methods are a powerful tool for investigating the low-energy behavior of interacting many-electron systems. They can be roughly divided into two categories: Real-space RG methods like the density matrix renormalization group (DMRG) \cite{dmrg} or contractor renormalization (CORE) \cite{morningstar} are based on finding the effective degrees of freedom in a relatively small local subsystem and then determine the non-local coupling to its environment, composed of the same type of subsystem. Momentum space methods, on the other hand, proceed by integrating out degrees of freedom with large kinetic energy, thus renormalizing the couplings of the degrees of freedom with low kinetic energy. The real space methods are better suited for problems where local correlations are strong, whereas momentum space methods are particularly useful when the interactions are moderate so that instabilities involve only a small region around the Fermi surface.

Our main interest here lies in Fermi surface instabilities, so that we focus on momentum space methods that allow to isolate the degrees of freedom around the Fermi surface. 
Whereas the momentum space RG does enable one to do this, it can not be used directly to obtain a low energy effective model of the many-fermion problem. This is due to the fact that instabilities of the Fermi surface manifest themselves in the form of divergences of the flowing coupling functions, which indicate the breakdown of the perturbative flow equations, that are valid when it is mainly the kinetic energy that determines the energy of a given state. In many cases, when the instability of the Fermi surface is driven by a single channel, the strong coupling problem at low energies can be tackled using mean-field approximations \cite{metznermeanfield}. In one dimension, bosonization has been applied successfully to a wide variety of problems (see e.g. \cite{giamarchi} and references therein). Useful as these methods are, they suffer from limitations regarding their range of applicability: The bosonization method is very powerful in one dimension, but has proven difficult to generalize to higher dimensions except when the Fermi liquid is stable \cite{fradkinboso}. The mean-field approach assumes long-range order, and can hence not describe quantum disordered phases. Given these limitations, we think a different approach to the problem may generate additional insights, and extend the range of systems that can be treated.

To this end, we propose to employ a basis transformation for the single electron states that allows to approximately map the strongly interacting system of fermions in the vicinity of the Fermi surface (subject to the renormalized couplings) onto a new lattice system where the interactions are short ranged in real space. As the coupling is comparable to the kinetic energy, real space methods are expected to give better results than the momentum space approach. More explicitly, we introduce an orthogonal set of wave packets with a characteristic width $\Delta x \sim M$ in real space, and a corresponding width $\Delta k\sim 2\pi/M$ in momentum space. By restricting the mean position and momentum of the wave packets to lie on a lattice, the system of wave packets can be made orthogonal and translationally invariant with period $2M$, twice the width of a wave packet. Because of their spatial extent, the wave packets average over large regions of real space, so that non-local interactions become much more localized in the new basis. From a technical point of view, this allows to use the real space cluster methods to find the effective degrees of freedom for long-wavelength physics and thus to derive an effective model for the problem at hand. From a more physical point of view, most Fermi surface instabilities are accompanied by binding of fermion pairs (particle-particle or particle-hole) and condensation of the bound pairs. This implies that for length scales less than the pair size, fermionic degrees of freedom provide an adequate description of the system. At larger length scales, however, the fermionic degrees of freedom are pushed to higher energies, and the low energy sector is described in terms of paired fermions, which have a bosonic character. Hence the fermion pairs at length scale $M$ may be approximated by local bosonic states in the wave packet basis. By integrating out the fermionic states at this scale, we derive an effective bosonic model for the larger length scales. This approach is similar in spirit to the derivation of bosonic effective actions by means of the Hubbard-Stratonovich transformation \cite{hubbard,stratonovich}. The main difference is that our approach is based on Hamiltonians instead of actions, which allows to use numerical methods such as exact diagonalization.

Wave packet bases are widely used in signal analysis and processing, and a plethora of bases with different properties exist (for a review, see e.g.~\cite{wavelets}). However, it turns out that it is not easy to achieve good localization in both momentum- and real-space with orthonormal bases of wave packets, so that for signal processing usually orthonormality is abandoned. For quantum mechanical calculations, orthonormality is crucial in order to preserve the fermionic anti-commutation relations. An ingenious way to preserve both features was pioneered by Wilson \cite{wilsonbasis} and later formalized in \cite{daubechies, discretewilson}.

This paper is the first part of a series in which we aim to apply the Wilson-Wannier basis to interacting fermion systems on a lattice. In this part we present the fundamentals of the approach. In order to keep the discussion transparent, we apply the method to a relatively simple system, the Hubbard chain at half-filling with weak repulsive interaction. Clearly, the physics of this model has already been discussed comprehensively in the literature (see e.g. \cite{giamarchi} and references therein), so that no new results can be expected. The current setup is not meant to compete with existing methods such as Bethe ansatz, bosonization or the density renormalization group \cite{dmrg}. Instead, we see the main advantage of our approach in the fact that it can be easily generalized to higher dimensions, which is not true of the methods that are specialized to one dimension.

%The logic of our approach is schematically shown in fig. \ref{fig:flowchart}. First, the degrees of freedom that are far away from the Fermi points are integrated out using the one-loop renormalization group. The main results of the RG flows are well-known, and we review them in sec. \ref{sec:rg}. 
\begin{figure}
\centering
\includegraphics[height=8cm]{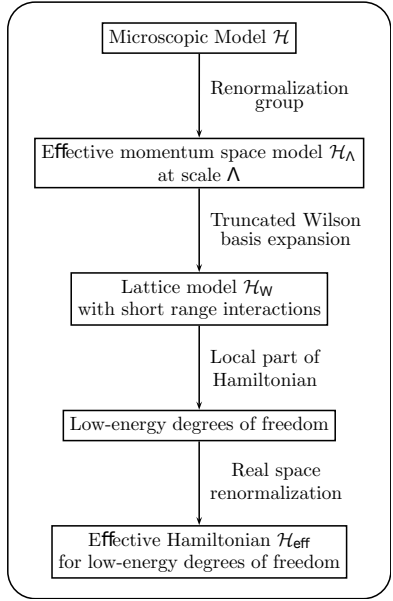}
\caption{Schematic algorithm of the wave packet method in an RG framework}
\label{fig:flowchart}
\end{figure}

The remainder of the paper develops the steps outlined in Fig.~\ref{fig:flowchart}. Sec.~\ref{sec:intro_phase_space} introduces the concept of phase space representations in an informal way, and relates its usefulness to the presence of scales that separate different regimes. The ideas are formalized in following section, where we briefly review an overcomplete wave packet basis that underlies the orthogonal Wilson-Wannier basis. The construction of the latter is discussed in Sec.~\ref{sec:wwbasis}. In general, the evaluation of matrix elements of operators in the WW basis has to be done numerically. In order to facilitate computations, in Sec.~\ref{sec:wwtrafo_1} we derive a systematic approximation scheme that can be used to obtain matrix elements as well as gain a more intuitive understanding of the results. The connection with correlations in fermion systems is established in Sec.~\ref{sec:wwpairing}, where we focus in particular on the manifestation of short range order in the two-particle density matrix. We show that the WW basis is well suited to capture the relevant correlations efficiently. Finally, we put these considerations to use in Sec.~\ref{sec:rg} by combining the WW basis with an RG calculation for the half-filled Hubbard chain. 

Since this paper is the first in a series of investigations, we conclude with an outlook of the application of the methods developed here to more general problems, and discuss improvements of the approximations made in the present work.

\section{Phase space representations and wave packet bases}
\label{sec:intro_phase_space}

\begin{figure}
\centering
\subfigure[]{\includegraphics[width=6cm]{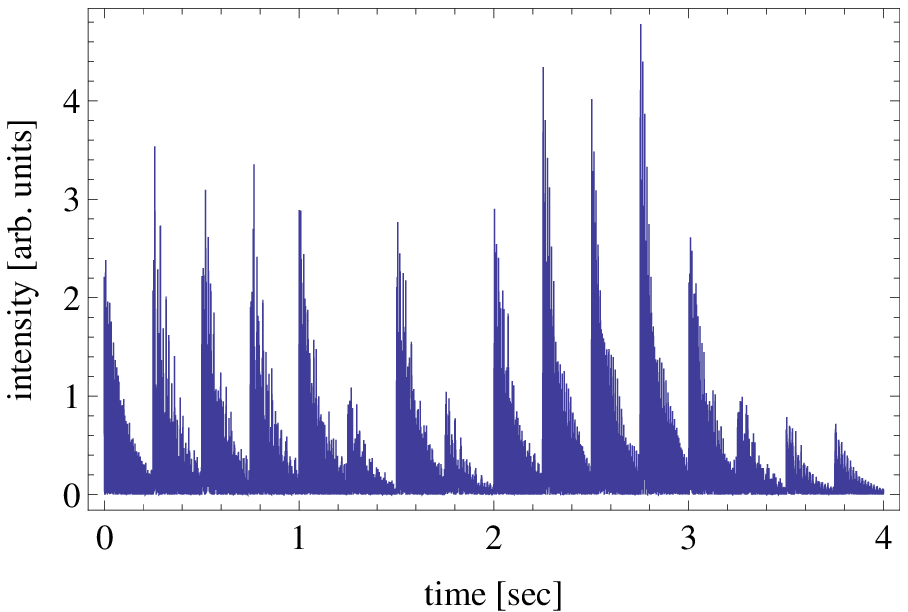} }
\subfigure[]{\includegraphics[width=6cm]{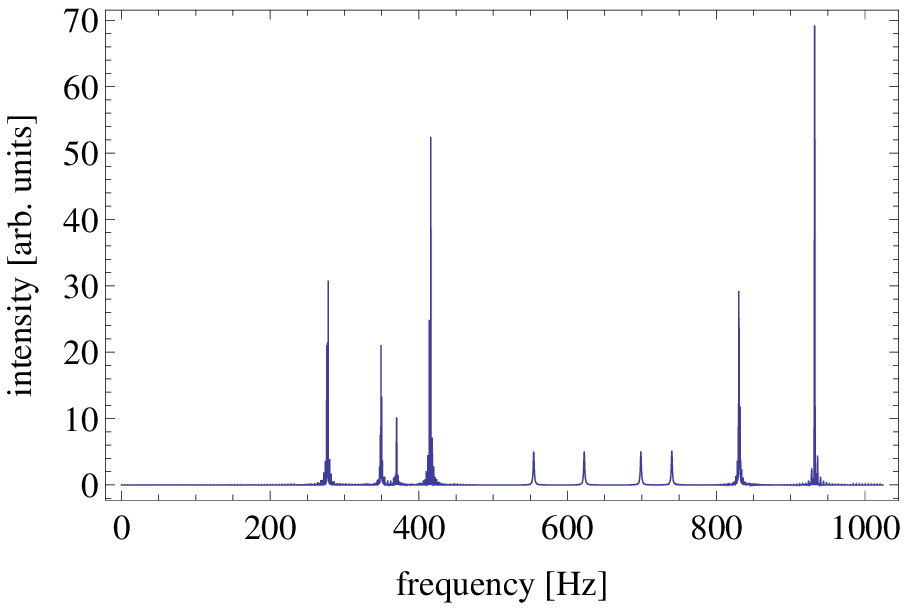}}
\caption{A short piece of music in the (a) time domain and (b) frequency domain.}
\label{fig:music}
\end{figure}

In this section we introduce the phase space representation of functions on one-dimensional lattices by means of overcomplete wave packet bases. These bases are commonly used in signal processing (see e.g.~\cite{wavelets}), where they allow to represent a time-dependent signal by a set of coefficients that are related to the signal strength around a grid of points in the time-frequency plane (i.e. the phase space). Mathematically, these coefficients are obtained by integrating the signal against a set of time and frequency translates of a single window function. While the phase space representation of a signal is computationally and mathematically more demanding than the time or frequency representation, its main advantage is that the structure of many signals is better represented in phase space than in either the frequency or the time domain, allowing for efficient signal compression as well as feature extraction \cite{wavelets}. As a basic example, Fig.~\ref{fig:music} shows the intensity distribution of a short piece of music in the time and frequency domains. In the time domain, it is easy to extract the beginning of each note (contained in the amplitude), but much harder to obtain its pitch (contained in the phase). Conversely, the frequency domain gives information about which notes are used (contained in the amplitude), but their temporal position is concealed in the phase. 

\begin{figure}
\centering
\includegraphics[width=6cm]{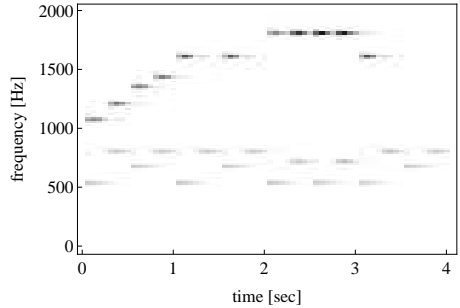}
\caption{Phase space representation of the signal shown in fig.~\ref{fig:music}. As opposed to the time and frequency representations, the structure of the signal is clearly visible: Each peak corresponds to a single note, both the temporal sequence of notes and their pitches are directly visible.}
\label{fig:music_phase_space}
\end{figure}

In order to arrive at a more descriptive representation, we observe that the time and frequency domain representations are not optimal because the signal contains structure on different time scales: The shortest scale is the sampling rate (here 2028Hz), corresponding to a time scale of $T_{\text{sample}} \approx 0.001$s. Second, the inverse frequency of a note is $T_{\text{pitch}}\lesssim 0.01$s, whereas a typical duration is $T_{\text{shape}}\gtrsim 0.1$s. Since individual notes are similar to plane waves on short time scales $\gtrsim T_{\text{pitch}}$, the signal is best represented in the frequency domain. On larger time scales $\gtrsim T_{\text{shape}}$, on the other hand, the temporal sequence of notes is better represented in the time domain. These findings suggest to cut the signal into pieces of duration $\lesssim T_{\text{shape}}$, and to represent each of these pieces (or windows) in the frequency domain. The position of the moving window yields a coarse-grained time coordinate, the Fourier components of the signal within the window yield a frequency coordinate. The dependence on these two coordinates resembles the classical phase space. We defer the mathematical details of this \emph{phase space representation} to the next section. The phase-space representation maps the one-dimensional signal onto the two-dimensional time-frequency plane (or phase-space), resulting in a plot like the one shown in Fig.~\ref{fig:music_phase_space}. It is evident that this representation is more efficient than both the time or frequency representation in capturing the two main pieces of information, namely the timing and pitch of the individual notes. 

In the following we argue that a similar simplification can be achieved in the analysis of some correlated electron systems as long as the interactions are not too strong. We consider the one-dimensional case only, where the correspondence to the above example is
\begin{center}
\begin{tabular}{ccc}
time & $\leftrightarrow$ & space \\
frequency & $\leftrightarrow$ & wave vector.
\end{tabular} 
\end{center}

In order to elucidate the length scales corresponding to the three time scales above, consider the case of a half-filled Hubbard chain with repulsive interactions. The length scale analogous to the inverse sampling rate $T_{\text{sample}}$ is the lattice spacing, which we set to unity in the following. It is well known (see e.g.~\cite{giamarchi}) that at low temperatures fermionic excitations are gapped, so that the equal time correlation function for single fermions decays exponentially on some length scale $\xi$ that depends on the interaction strength. On larger length scales, the physics is determined by the spin degrees of freedom which remain gapless, similar to the antiferromagnetic Heisenberg model. At length scales $< \xi$, the fermions are expected to behave similar to the noninteracting case, since the kinetic energy dominates on these short length scales. In analogy to the example above, the system is easiest to describe in momentum space on these scales. By contrast, at scales larger than $\xi$, individual fermions are confined and the physics is better described in terms of localized (on scale $\xi$) spins, so that a real space representation is more adequate. Motivated by these heuristic considerations, we introduce the phase space description in terms of tight frames in the next section.

\section{Mathematics of phase space representations for one-dimensional lattices}
\label{sec:mathematics_phase_space}

In this section we introduce the mathematics behind phase space representations for one-dimensional lattices. We begin with a short review of a particular kind \emph{tight frame} (to be defined below), an overcomplete basis that can be used to describe the phase space density of functions on the lattice. This basis is similar to the coherent state representation in quantum mechanics, in that the basis functions are generated by shifting a single window function in real and momentum space. This representation allows to analyze the phase space content of correlation functions, which may be useful by itself. At the same time, however, the overcompleteness makes it difficult to do actual computations. Hence we proceed by introducing a trick (found in \cite{wilsonbasis} and formalized in \cite{daubechies}) that allows to obtain an orthogonal basis from the overcomplete representation that inherits its key advantages, the so-called Wilson or Wilson-Wannier (WW) basis. 

\subsection{Phase space representation}

In the following we specialize to the case of a one-dimensional lattice with $N$ sites and periodic boundary conditions. The individual lattice sites are labelled by the index $j=0,\ldots,N-1$. We define the phase space corresponding to this lattice to be the two-dimensional lattice consisting of the points $(i, 2\pi/N j)$, where $i,j = 0,\ldots N-1$, so that the phase space consists of $N^2$ points in total.

We want to construct a basis with basis functions that are localized around the points of a rectangular lattice in phase space. We generate the basis from a single window function $g(j)$. We demand that both $g(j)$ and its Fourier transform $\tilde{g}(p)$ are localized and symmetric around zero (e.g.~a Gaussian), and that $g(j)$ is normalized. The basis is generated by applying two operations to the window function $g(j)$: 
\begin{enumerate}
\item[i)] Shifts in real space, $g(j) \rightarrow g\left(j- M m\right)$, where $M$ is an integer that divides $N$, and $m=0,\ldots,N/M -1$ is an integer mod $N/M$.  
\vspace{.1cm}
\item[ii)] Modulation, $g(j) \rightarrow e^{i K k j}$, where $K/2\pi$ is an integer that divides $N$ and $k = 0,\ldots 2\pi N K/-1$ is an integer mod $2\pi N/K$.
\end{enumerate}
Note that the two operations do not commute in general (unless $MK = 2\pi$), so that an ordering has to be specified. The basis states $\ket{g_{mk}}$ are defined  as
\begin{eqnarray}
\braket{j}{g_{mk}} &=& g_{mk}(j), \label{eq:def_wave_packet_frame} \\
g_{mk}(j) &=& e^{i K k j} g\left(j-Mm\right),\label{eq:def_g_mk}
\end{eqnarray}
where $g_{mk}(j)$ has mean position $Mm$ and mean momentum $K k$. The lattice spacings are $M$ in the real space direction and $K$ in the momentum space direction. The number of states in the basis is determined by number of phase space points per unit cell of the phase space lattice, 
\begin{eqnarray}
B &=& \frac{\text{number of points in phase space}}{\text{number of points per unit cell}}\nonumber \\
&=& \frac{2\pi N}{M K}.
\label{eq:tight_frame_number_of_states}
\end{eqnarray}
In particular, $B=N$ if $MK = 2\pi$, in which case the $g_{mk}(j)$ form a complete (non-orthogonal) basis. A particularly simple resolution of the identity is obtained if the window function satisfies 
\begin{equation}
\sum_{m=0}^{N/M-1} g\left(j - m M\right) g\left(j - M\left(m+ 2 l\right)\right) = \frac{1}{M}\delta_{l,0}
\label{eq:gconditions}
\end{equation}
for all integers $j$ and $l$. In this case, it has been proven in \cite{daubechies, discretewilson} that
\begin{equation}
\sum_{mk} \ket{g_{mk}}\bra{g_{mk}} = \frac{MK}{2\pi}.
\label{eq:tight_frame_resolution_of_identity}
\end{equation}
The advantage of this resolution of the identity is that wave packet expansions preserve the norm of the expanded vector. Moreover, it enables the construction of an orthonormal basis from the overcomplete basis $\ket{g_{mk}}$, that we discuss in Sec.~\ref{sec:wwbasis}.

\subsection{Analytical window functions}

In general, window functions that satisfy (\ref{eq:gconditions}) have to be constructed numerically. However, a special class of window functions can be readily constructed analytically. The key condition that has to be imposed is that the window function have compact support in either real or momentum space. Here we focus on the latter case. To this end, we introduce the Fourier transform $\tilde{g}(p)$ of $g(j)$:
\begin{equation}
\tilde{g}(p) = \frac{1}{\sqrt{N}} \sum_j e^{-i p j} g(j).
\end{equation}
Then we demand that $\tilde{g}(p)$ has compact support:
\begin{equation}
\tilde{g}(p) = 0 \text{ for } |p| \geq K.
\label{eq:condition_p_local}
\end{equation}
Condition (\ref{eq:condition_p_local}) states that only shifted window functions that are nearest neighbors in momentum space overlap, i.e.
\begin{equation}
\tilde{g}_{mk}(p)\,\tilde{g}_{m'k'}(p) = 0 \; \text{ for } |k-k'| > 1. 
\end{equation}
From condition (\ref{eq:condition_p_local}) one sees that the number of parameters needed to fix $\tilde{g}(p)$ is $N/2M$. For a band limited window function the conditions (\ref{eq:gconditions}) become \cite{daubechies, discretewilson, ossadnikthesis}
\begin{equation}
\left|\tilde{g}\left(p\right)\right|^2 + \left|\tilde{g}\left(K - p \right)\right|^2 = \frac{2M}{N} \;\text{ for } 0 \leq p \leq K.
\label{eq:orthogonality_condition_p_loc}
\end{equation}

This implies that the values 
\begin{eqnarray}
\tilde{g}(0) &=& \sqrt{\frac{N}{2M}}, \nonumber\\
\tilde{g}\left(K/2\right) &=& \frac{1}{2}\sqrt{\frac{N}{M}}
\end{eqnarray}
are fixed. For the remaining momenta, any value $\tilde{g}(p) \leq \sqrt{2M/N}$ can be chosen for $0<p<K/2$, the remaining values are fixed by (\ref{eq:orthogonality_condition_p_loc}) and (\ref{eq:condition_p_local}), and $\tilde{g}(p)=\tilde{g}(-p)$.

Window functions that satisfy (\ref{eq:condition_p_local}) are listed in Tab.~\ref{tab:windowfunction} for the cases $N/M=2,4$. Note that for $N/M=2,4$, the window function is unique, whereas for $N/M > 4$ it is not. 

\begin{table}
\centering
\begin{tabular}{| c | c | c | c | c |}
\hline 
$N/M$ & $\tilde{g}\left(0\right)$ & $\tilde{g}\left(\frac{2\pi}{N}\right)$    \\
\hline
2      &   1                                  &          0                                \\
\hline
4      &   $ \frac{1}{\sqrt{2}}$         &        $ \frac{1}{2}  $      \\ 
\hline
\end{tabular}
\label{tab:windowfunction}
\caption{Analytical window functions in momentum space for small lattices with $N/M=2,4$. The value of $\tilde{g}(p)$ for all other momenta is either zero or related to the ones given by symmetry.}
\end{table}

\subsection{Wave packet transformation}

The phase space representation can be used to decompose arbitrary $\ket{f}$ by inserting the resolution of identity (\ref{eq:tight_frame_resolution_of_identity}), 
\begin{eqnarray}
\ket{f} &=& \frac{2\pi}{MK} \sum_{mk} \ket{g_{mk}} \braket{g_{mk}}{f}\nonumber \\
&\equiv& \frac{2\pi}{MK} \sum_{mk} \bar{f}_{mk} \, \ket{g_{mk}}, 
\label{eq:wave_packet_insert_resolution}
\end{eqnarray}
where
\begin{eqnarray}
\bar{f}_{mk} &=& \braket{g_{mk}}{f} \label{eq:wave_packet_transform} \\
&=&\sum_j g^\ast_{mk}(j) \, f(j) \nonumber \\
&=& \sum_p \tilde{g}^\ast_{mk}(p) \, \tilde{f}(p). \nonumber
\end{eqnarray}
The coefficients $\bar{f}_{mk}$ capture the weight of $f(j)$ in different parts of the phase space. In the following, we will refer to the transformation (\ref{eq:wave_packet_transform}) as the \emph{wave packet transformation}, and to the coefficients $\bar{f}_{mk}$ as the \emph{wave packet transform} of the function $f(j)$ (or its Fourier transform $\tilde{f}(p)$). $\bar{f}_{mk}$ can be used to define the phase space density $\left|\bar{f}_{mk}\right|^2$ of a function, which is the density that is plotted above in Fig.~\ref{fig:music_phase_space}.

In a similar way, the wave packet transform of matrices and higher ranked tensors is obtained by applying (\ref{eq:wave_packet_insert_resolution}) to each index. In quantum mechanics, tensor indices may correspond to fermion annihilation or creation, and the conjugate version of (\ref{eq:wave_packet_insert_resolution}) is needed in the latter case. For example the wave packet transform $\bar{t}_{mk,m'k'}$ of the hopping matrix $t(j,j')$ can be obtained from 
\begin{equation}
\begin{split}
t(j,j') =& \bra{j} \hat{t} \ket{j'} \nonumber \\
=& \left(\frac{2\pi}{MK}\right)^2\sum_{mk,m'k'}  \bra{g_{mk}} \hat{t} \ket{g_{m'k'}}\\ & \times \braket{j}{g_{mk}}\braket{g_{m'k'}}{j'} \\
=& \left(\frac{2\pi}{MK}\right)^2\sum_{mk,m'k'} g_{mk}(j)\,g^\ast_{m'k'}(j')  \bar{t}_{mk;\,m'k'}
\end{split}
\end{equation}
It is given by
\begin{eqnarray}
\bar{t}_{mk;\,m'k'} &=& \sum_{j,j'} g_{mk}^\ast(j) \, t(j,j') \, g_{m'k'}(j')\nonumber \\
&=& \sum_{p,p'} \tilde{g}_{mk}^\ast(p) \tilde{t}(p,p') \tilde{g}_{m'k'}(p'),
\label{eq:wp_trafo_hopping}
\end{eqnarray}
where the second line is the momentum space version of the first line. 

The wave packet transform of $\ket{f}$ separates slow and fast parts: The behavior at distances shorter than $M$ are encoded in the momentum part $k$ of the wave packet, whereas slow variations are contained in the real space part, $m$. In order to obtain a good representation of $f(j)$, the parameter $M$ has to be adjusted to the characteristics of the system. In the music example above, $M \approx 0.05$s is a reasonable choice because then $T_{\text{pitch}} < M < T_{\text{shape}}$, so that information about the pitch is contained (mainly) in $k$, and the position and shape of the different notes is (mainly) contained in $m$. For the fermion pairing problem, $M \approx \xi$ is the most natural choice: At distances much less than $\xi$ the kinetic energy dominates, so that a momentum space representation is preferable, for large length scales fermions occur in pairs only, so that a real space description is more adequate.

\section{Wilson-Wannier basis in one dimension}
\label{sec:wwbasis}

The phase space representation for one-dimensional lattices introduced above has the advantage that the interpretation of the coefficients of the wave packet transform (\ref{eq:wave_packet_transform}) is relatively easy to evaluate and interpret. However, for quantum mechanical applications it is better to work with an orthonormal basis, so that the canonical anti-commutation relations are preserved. Following \cite{wilsonbasis,daubechies,discretewilson}, we now construct the Wilson-Wannier (WW) basis from the phase space representation above for the case $MK = \pi$.

According to (\ref{eq:tight_frame_number_of_states}), this phase space representation consists of $2N$ states. Hence the number of states has to be reduced by a factor of two in order to obtain a complete basis. The prescription that yields an orthonormal basis \cite{wilsonbasis, daubechies} is to divide the phase space lattice $(M m, K k)$ into an even and an odd sublattice, dependent on the parity of $m+k$. Then states at even (odd) phase space lattice points are projected to even (odd) symmetry around the center of the wave packet. This procedure eliminates half of the states, and the resulting basis is orthogonal if (\ref{eq:gconditions}) is satisfied \cite{daubechies, discretewilson}. Denoting the WW basis states by $\ket{mk}$, where $m=0,\ldots,N/M-1$ and $k=0,\ldots,M$, their relation to the phase space representation is thus
\begin{eqnarray}
\ket{mk} &\propto& \ket{g_{m,k}} + \left(-1\right)^{m+k} \ket{g_{m,-k}} \nonumber \\
&=& \sum_{\alpha=\pm 1} \alpha^{m+k} \ket{g_{m,\alpha k}}.
\end{eqnarray}
We use the remaining freedom in the  choice of the prefactor to normalize the states, and to make all the $\psi_{mk}(j) = \braket{j}{mk}$ real. It is easy to verify that this is achieved by
\begin{equation}
\ket{mk} =  \frac{1}{\sqrt{2 H_k}}\sum_{\alpha=\pm 1} e^{-i\alpha\phi_{m+k}} \ket{g_{m,\alpha k}},
\label{eq:defpsi}
\end{equation}
where 
\begin{equation}
\phi_{m+k} = \left\{\begin{array}{lcc} 0 & \text{for} & m+k \text{ even} \\ \frac{\pi}{2} & \text{for} & m+k \text{ odd}\end{array}\right.,
\label{eq:def_phi}
\end{equation}
and 
\begin{equation}
H_k = \left\{\begin{array}{lcl} 2 & \text{for} & k = 0, M \\ 1 & \text{for} & k=1,\ldots,M-1\end{array}\right. .
\end{equation}
The wave functions $\psi_{mk}(j)$ are thus given by
\begin{eqnarray}
\psi_{mk}(j) &=& \frac{1}{\sqrt{2H_k}}\sum_{\alpha=\pm 1}e^{-i\alpha \phi_{m+k}} g_{m,\alpha k}(j)\\
&=& \frac{1}{\sqrt{2H_k}}\left\{\begin{array}{c} 2 \cos K k j \\ 2i \sin K k j\end{array}\right\} \,g(j-Mm),\nonumber
\end{eqnarray}
where the cos (sin) is used for even (odd) $m+k$. The WW expansion of a state $\ket{f}$ can be conveniently expressed using the wave packet transform $\bar{f}_{mk}$,
\begin{eqnarray}
\ket{f} &=& \sum_{mk} \ket{mk} \braket{mk}{f}  \\
&=& \sum_{mk} \left[\frac{1}{\sqrt{2 H_k}} \sum_\alpha e^{-i\alpha \phi_{m+k}} \bar{f}_{m,\alpha k}\right]\, \ket{mk}\nonumber
\end{eqnarray}

\begin{figure}
\centering
\includegraphics[width=6cm]{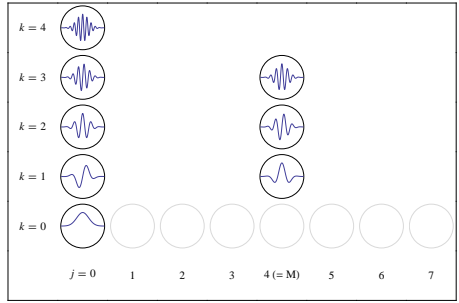}
\caption{Schematic representation of the relation of Wilson-Wannier functions to the real space lattice. The figure shows one unit cell of the Wilson basis for $M=4$. $j$ labels lattice sites in the real lattice, and gray circles represent these sites. The WW momentum is denoted by $k$ and runs in the vertical direction. The $2M = 8$ sites in the original lattice are replaced by two sets of states centered around $j=0$ and $j=M$ in the Wilson basis. Note that the two superlattice sites within one unit cell are inequivalent, which can be seen best from the fact that the states $k=0$ and $k=M$ exist only once per unit cell.}
\label{fig:wilson_basis}
\end{figure}

\begin{figure}
\centering
\includegraphics[width=6cm]{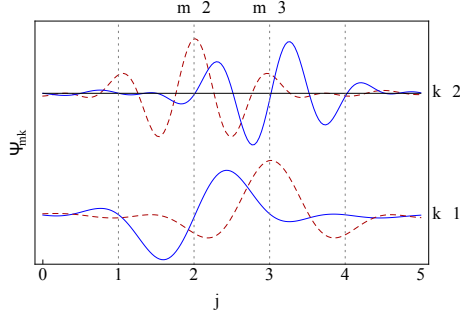}
\caption{A subset of the Wilson-Wannier basis functions within one unit cell of the basis for $N=2^{10}$ and $M=2^5$. The figure shows $\psi_{mk}(j)$ with $m=2,3$ and $k=1,2$. States with $k=2$ are offset vertically for sake of clarity. The centers of the wave packets in real space are marked by the dotted gray lines at positions $j = M m$. The parity of the states is given by $(-1)^{m+k}$ and hence follows a checkerboard pattern in the $m-k$-plane. States with even parity are shown in red (dashed line), states with odd parity in blue (solid line).}
\label{fig:wwbasisfunctions}
\end{figure}

The unit cell for the basis functions is $2M$ because of the phase factors $e^{\pm i \phi_{m+k}}$ in (\ref{eq:defpsi}) that are different on adjacent WW sites but identical on second nearest neighbor sites. The states with $k=0,M$ appear only once per unit cell, for all other $k$ there are two states per unit cell for even and odd parity. A schematic picture of the basis function in one unit cell is shown in Fig.~\ref{fig:wilson_basis}. There are $N/2M$ unit cells in total. Note that this figure is intended to show how the states are rearranged in the new basis only, and that it does not reproduce the shape of the wave packets correctly. The real space form of the wave packets within one unit cell is shown in Fig.~\ref{fig:wwbasisfunctions}. The figure shows wave packets with $m=2,3$ and $k=1,2$. The parity of the states follows a checkerboard pattern in the $m-k$-plane, where nearest neighbors always have opposite parity.

The fermion creation operator $\gamma^\dagger_{mk}$ that creates a fermion in the state $\ket{mk}$ is given by
\begin{eqnarray}
\gamma_{mk}^\dagger &=& \sum_j \psi_{mk}(j) \, c^\dagger_j \label{eq:c_j_to_gamma}\\
&=& \sum_p \tilde{\psi}^\ast_{mk}(p) \, \tilde{c}^\dagger_{p}.\label{eq:c_p_to_gamma}
\end{eqnarray}
Eqns.~(\ref{eq:c_j_to_gamma}, \ref{eq:c_p_to_gamma}) can be used to transform any many-fermion operator into the WW basis. Similar to the transformation of a single particle state, the WW representation of many-body operators can be expressed using the wave packet transform, Eq.~(\ref{eq:wave_packet_transform}). We continue with the example of the hopping matrix from above. The kinetic energy part of the Hamiltonian is $\ham{kin} = \sum_p \epsilon(p)\,\tilde{c}_p^\dagger\,\tilde{c}^\pdag_p$, so that according to (\ref{eq:c_p_to_gamma}) its WW representation is
\begin{eqnarray}
\ham{kin} &=& \sum_{mk,m'k'}t_{mk,m'k'}\,\gamma^\dagger_{mk}\,\gamma^\pdag_{m'k'},
\end{eqnarray}
where
\begin{widetext}
\begin{eqnarray}
 t_{mk,m'k'}&=& \sum_p \epsilon(p) \tilde{\psi}_{mk}^\ast(p) \, \tilde{\psi}_{m'k'}(p)\nonumber \\
&=& \frac{1}{2 \sqrt{H_k\,H_{k'}}} \sum_{\alpha,\alpha'} e^{-i\left(\alpha\phi_{m+k}+\alpha'\phi_{m'+k'} \right)}\sum_p \epsilon(p)\,\tilde{g}_{m,\alpha k}^\ast(p)\,\tilde{g}_{m',\alpha' k'}(p) \nonumber \\
&=&\frac{1}{2 \sqrt{H_k\,H_{k'}}} \sum_{\alpha,\alpha'} e^{-i\left(\alpha\phi_{m+k}+\alpha'\phi_{m'+k'} \right)} \bar{t}_{m,\alpha k;\;m',\alpha' k'}.
\label{eq:wp_to_ww_hopping}
\end{eqnarray} 
\end{widetext}
In the last line we have inserted the wave packet transform $\bar{t}_{mk,m'k'}$ of the hopping matrix from Eq.~(\ref{eq:wp_trafo_hopping}).

To conclude this section, Tab.~\ref{tab:variables_1d} summarizes the meaning of the symbols introduced above for later reference.

\begin{widetext}
\begin{table}
\begin{center}
\begin{tabular}{| c | c | p{5cm} |}
\hline 
symbol & range & meaning \\
\hline
$j$ & $0,\ldots, N$ & position in real space \\ 
\hline 
$p$ & $-\pi+\frac{2\pi}{N}, -\pi+2\frac{2\pi}{N},\ldots, \pi-\frac{2\pi}{N}, \pi$ & momentum \\
\hline
$m$ & $0,\ldots, N/M-1$ & WW position label for state with center position $\bar{j}=Mm$ \\
\hline
$k$ & $0,\ldots, M$ & WW momentum label for state with center momenta $\bar{p}=\pm K k$\\
\hline
$\alpha$ & $\pm 1$ & Sign of $k$ in the definition of WW basis states \\
\hline
$M$ & even & Real space shift length. Size of WW unit cell is $2M$ \\ 
\hline
$K$ & $\frac{\pi}{M}$ & Momentum shift length \\
\hline
$\ket{g_{mk}}$ & $m=0,\ldots,N/M-1$;  $k=-M+1,\ldots,M$ & Wave packet state \\
\hline
$\ket{mk}$ & $m=0,\ldots,N/M-1$;  $k=0,\ldots,M-1$ & Wilson basis state \\
\hline
%$L$ & $\frac{N}{M}$ & Range of $m$ \\
%\hline
\end{tabular}
\end{center}
\caption{List of symbols for the description of the spatial degrees of freedom.}
\label{tab:variables_1d}
\end{table}
\end{widetext}

\section{WW representation of hopping and interaction operators}
\label{sec:wwtrafo_1}

In this section we begin to apply the WW basis to one-dimensional fermion systems by discussing the WW representation of the hopping and interaction operators. The purpose is twofold: First, we show that for short ranged interactions (compared to $M$), the transformation can be simplified by means of a gradient expansion in momentum space. This method reduces the computational effort to the evaluation of few convolutions involving the window function. Second, we relate the resulting matrix elements to the dynamics of wave packet states which leads to a more intuitive grasp of them. For sake of simplicity we suppress spin indices throughout this section.

\subsection{Hopping}
\label{sec:ww_trafo_hopping_1}

When the wave packet size $M$ is greater than the range of the interaction described by the tensor, the transformation can be simplified by employing a gradient in expansion as follows: In general, for a translationally invariant system, the interaction tensors conserve the momentum, so that one argument is fixed by a delta function. The remaining momenta describe the dependence on the relative positions of the particles. When the interaction is short ranged, the latter part is slowly varying in momentum space compared to the width of $\tilde{g}(p)\sim 2K = 2\pi/M$. As a consequence, one can Taylor expand the momentum dependence. The point around which one expands depends on whether or not the coarse grained momenta $k$ are conserved (i.e. $k=k' \mod 2M$ in the above example). When they are conserved, we can expand around $p=Kk,p'=K k'$, otherwise one has to expand around a nearby point where momentum conservation is satisfied. We treat the former case only, but the generalization does not introduce complications. Setting $\tilde{t}(p,p') = \epsilon(p) \delta(p-p')$, we can then expand around $p=Kk$
\begin{equation}
\begin{split}
\bar{t}_{mk,m'k} &= \sum_{p} \epsilon(p)\,\tilde{g}_{mk}^\ast(p)\, \tilde{g}_{m'k}(p) \\
%&=& \sum_{p} \left[\epsilon(Kk) + \epsilon'(Kk) \left(p-Kk\right) + \ldots \right] \tilde{g}^\ast_{mk}(p)\tilde{g}_{m'k}(p)\nonumber \\
&= \sum_{n=0}^\infty \frac{1}{n!}\epsilon^{(n)}(Kk) \left[\sum_p p^n\,\tilde{g}^\ast_{m,0}(p)\tilde{g}_{m',0}(p)\right] 
\label{eq:wave_packet_gradient_expansion}
\end{split}
\end{equation}
where $\epsilon^{(n)}(Kk)$ is the $n$-th derivative of $\epsilon(p)$ evaluated at $p=Kk$. We have used the definition (\ref{eq:def_g_mk}) of the shifted window function $g_{mk}(j)$ in order to shift $p$ in the last line. Dimensional analysis reveals that each power of $p$ in the expansion (\ref{eq:wave_packet_gradient_expansion}) contributes an additional power of $1/M$ in the result of the summation, so that only a few terms are needed when interactions are short ranged. As a consequence, the computational effort for the transformation is dramatically reduced, since only a handful of moments of products of the window functions have to be evaluated.

Now we obtain an analytical approximation of the WW representation of the hopping matrix, using the connection (\ref{eq:wp_to_ww_hopping}) between wave packet transformation and WW basis, and the analytical window function with $N/M=4$ (see Tab.~\ref{tab:windowfunction}). We keep terms up to $O\left(1/M\right)$, which yields
\begin{widetext}
\begin{eqnarray}
\ham{kin} &\approx &\sum_k \sum_{m,m'} \gamma^\dagger_{mk}\,\gamma^\pdag_{m'k} \Big[ \epsilon\left(K k\right) \delta_{m,m'} + \left(-1\right)^{m} \frac{\pi}{4M} \epsilon'\left(Kk\right) \delta_{m+1,m'} \Big] \; + \; \text{ h.c.}
\label{eq:ham_kin_ww_1}
\end{eqnarray}
\end{widetext}
There are two approximations used in obtaining (\ref{eq:ham_kin_ww_1}): The gradient expansion and the analytical approximation of the window function. The gradient expansion to order $1/M$ splits the hopping operator into two terms. The diagonal part is determined by the mean kinetic energy of a wave packet, $\bra{g_{mk}}\ham{kin} \ket{g_{mk}} = \epsilon(Kk) + O(1/M^2)$. The second term describes the propagation of a wave packet with the group velocity, according to
\begin{eqnarray}
\text{hopping rate} &\approx& \frac{\text{group velocity}}{\text{distance}}\nonumber \\
&=& \frac{\epsilon'}{M},
\end{eqnarray}
with a prefactor $\pi/4$ of order unity. 

The approximate window function shows up in the evaluation of convolutions of the window function, such as $\sum_p \tilde{g}^\ast_{mk}(p) \tilde{g}_{m'k'}(p)$ in (\ref{eq:wave_packet_gradient_expansion}). This leads to conservation of $k$ and truncation of the hopping range to $|m'-m|=1$ (instead of rapid decay for larger distances). As long as $Kk$ is not close to the band edges, $\epsilon'(Kk) \gg \epsilon''(Kk)/M$  for large enough $M$, so that the first approximation is justified. The second approximation introduces larger errors that do not vanish systemically for large $M$. However, we emphasize that it is not difficult to improve the approximation, and the main reason that it is used here is that it yields compact and analytical results for the hopping matrix elements.

\subsection{Interaction}
\label{sec:ww_trafo_interaction_1}

Now we turn to the WW representation of two-body interactions. In order to treat general interactions, we include the spin depedence from now on. We parametrize the general translationally invariant interaction in momentum space as
\begin{equation}
\begin{split}
\ham{int} =& \frac{1}{2N} \sum_{p_1\cdots p_4}  \tilde{J}\left(p_1,p_3\right) \, \tilde{J}\left(p_2,p_4\right)\\
& \; \times \delta\left(p_1+p_2-p_3-p_4\right)\,\tilde{U}\left(p_1,p_2;\; p_3,p_4\right)
\end{split}
\end{equation}
where
\begin{equation}
\tilde{J}\left(p_1,p_2\right) = \sum_s \tilde{c}^\dagger_{p_1}\,\tilde{c}^\pdag_{p_2}.
\end{equation}

The WW representation of the interaction follows directly from the application of the single-particle operator formula, Eq.~(\ref{eq:wp_to_ww_hopping}), to each $\tilde{J}(p_1,p_2)$ separately. We obtain
\begin{equation}
\begin{split}
\ham{int} =&\frac{1}{2} \sum_{m_1k_1,\ldots,m_4,k_4}U_{m_1k_1,m_2k_2;\;m_3k_3,m_4k_4}\\ & \; \times J_{m_1k_1,m_3k_3}\,J_{m_2k_2,m_4k_4},
\end{split}
\end{equation}
where
\begin{equation}
J_{m_1k_1,m_2k_2} = \sum_s \gamma^\dagger_{m_1k_1s}\,\gamma^\pdag_{m_2k_2s}.
\end{equation}

The transformed interaction $U_{m_1k_1,\ldots,m_4k_4}$ is given by
\begin{widetext}
\begin{equation}
\begin{split}
U_{m_1k_1,\ldots,m_4k_4} =& \frac{1}{N} \sum_{p_1\cdots p_4} \Big[\delta\left(p_1+p_2-p_3-p_4\right)\,\tilde{U}\left(p_1,\ldots,p_4\right) \,\tilde{\psi}_{m_1k_1}^\ast(p_1)\tilde{\psi}_{m_2k_2}^\ast(p_2)\tilde{\psi}_{m_3k_3}(p_3)\tilde{\psi}_{m_4k_4}(p_4)\Big] \\
=& \frac{1}{4\sqrt{H_{k_1}\cdots H_{k_4}}} \sum_{\alpha_1\cdots \alpha_4} e^{i\left(\alpha_1\phi_1+\alpha_2\phi_2-\alpha_3\phi_3-\alpha_4\phi_4\right)} \,\bar{U}_{m_1,\alpha_1k_1,\ldots,m_4,\alpha_4k_4},
\end{split}
\end{equation}
\end{widetext}
where we have used the wave packet transform
\begin{equation}
\begin{split}
\bar{U}_{m_1k_1,\ldots,m_4k_4} = &\frac{1}{N} \sum_{p_1\cdots p_4} \Big[\delta\left(p_1+p_2-p_3-p_4\right) \\ & \; \times \;\tilde{U}\left(p_1,\ldots,p_4\right) \\
& \; \times\;\tilde{g}_{m_1k_1}^\ast(p_1)\tilde{g}_{m_2k_2}^\ast(p_2)\\& \times \tilde{g}_{m_3k_3}(p_3)\tilde{g}_{m_4k_4}(p_4)\Big] 
\end{split}
\label{eq:wp_interaction}
\end{equation}
of the interaction $\tilde{U}(p_1,\ldots,p_4)$. Repeating the procedure from Sec.~\ref{sec:ww_trafo_hopping_1}, we apply the gradient expansion to (\ref{eq:wp_interaction}), focussing on matrix elements that conserve the WW momentum $k$, i.e. $k_1 + k_2 = k_3 + k_4 \mod 2M$ in (\ref{eq:wp_interaction}). The leading order term (of order $M^0$) is
\begin{equation}
\begin{split}
\bar{U}_{m_1k_1,\ldots,m_4k_4} &\approx \tilde{U}\left(Kk_1,\ldots,Kk_4\right)\, \sum_j \prod_{i=1}^4 g_{m_ik_i}(j).
\end{split}
\end{equation}
% longer version of above equation
%\begin{equation}
%\begin{split}
%\bar{U}_{m_1k_1,\ldots,m_4k_4} &\approx \frac{\tilde{U}\left(Kk_1,\ldots,Kk_4\right)}{N} \sum_{p_1\cdots p_4} \Big[\delta\left(p_1+p_2-p_3-p_4\right)\,\\&\hphantom{\frac{\tilde{U}\left(Kk_1,\ldots,Kk_4\right)}{N}\sum_{p_1\cdots p_4} }\times\tilde{g}_{m_1k_1}^\ast(p_1)\tilde{g}_{m_2k_2}^\ast(p_2) \tilde{g}_{m_3k_3}(p_3)\tilde{g}_{m_4k_4}(p_4)\Big]\\
%&=\tilde{U}\left(Kk_1,\ldots,Kk_4\right)\, \sum_j \prod_{i=1}^4 g_{m_ik_i}(j).
%\end{split}
%\end{equation}
Using the definition (\ref{eq:def_g_mk}) of the shifted window functions, this can be further simplified because the convolution of the four window functions depends on $\sum_i k_i$ only. For $k$-conserving matrix elements, we can then define
 \begin{eqnarray}
\bar{U}_{m_1k_1,\ldots,m_4k_4}&\approx& \tilde{U}\left(Kk_1,\ldots,Kk_4\right)\,,\label{eq:wp_interaction_simple}\\
&& V\left(m_1,\ldots,m_4\right) \nonumber \\
 V(m_1,\ldots,m_4) &=& \sum_j \prod_{i=1}^4 g\left(j-M m_i\right). \label{eq:def_V}
\end{eqnarray}
The $k$-dependence of $\bar{U}_{m_1k_1,\ldots,m_4k_4}$ thus reflects the (short-ranged) position dependence of the interaction. The $m$-dependence originates in the window function only and is independent of the interaction. The $m$-dependence of the transformed interaction is thus the same as for an onsite interaction in this approximation.
The final expression for the WW representation of the interaction to leading order in the gradient expansion is thus
\begin{widetext}
\begin{equation}
\begin{split}
U\left(m_1k_1,\ldots,m_4 k_4\right) \approx \frac{V(m_1,\ldots,m_4)}{4\sqrt{H_{k_1}\cdots H_{k_4}}} \, \sum_{\alpha_1\cdots \alpha_4} \tilde{U}(K\alpha_1k_1,\ldots,K\alpha_4k_4)  \times e^{i\left(\alpha_1\phi_1+\alpha_2\phi_2-\alpha_3\phi_3-\alpha_4\phi_4\right)},
\end{split}
\label{eq:U_ww_final_formula}
\end{equation}
\end{widetext}
which is the main result of this section.

Observing that (\ref{eq:wp_interaction_simple}) is just the density of a wave packet state (i.e. $1/M$) squared, we infer heuristically that 
\begin{equation}
\begin{split}
\text{interaction} &\propto  \text{density}^2 \times \text{wave packet size} \\
&\propto \frac{1 }{M},
\end{split}
\end{equation}
so that for large enough $M$ it is consistent with the treatment of the hopping operator above to keep the leading term of the gradient expansion only.

Finally, we compute values of $V(m_1,\ldots,m_4)$ for the most important cases using the analytical window functions from Tab.~\ref{tab:windowfunction}. We consider the cases $m_1=m_2=m_3=0, m_4=m$, where three operators reside on one site, and $m_1=m_2=0, m_3=m_4=m$, where two operators are located on the same site. Note that only the relative positions matter due to the residual translational invariance of the wave packet states. Approximate analytical values of $V\left(m_1,\ldots,m_4\right)$ for these cases are tabulated in Tab.~\ref{tab:V_values}. Interactions decay rapidly, with spatial separation, so that $V(0,0,0,2)/V(0,0,0,0) \approx 1/32$. Consequently, we will take only nearest neighbor interactions into account. The table also shows the corresponding value for the case that the window function for $N/M=2$ is used. Since these are very similar, but simpler, we will use the latter in the following.

\begin{table}
\centering
\begin{tabular}{| c |  c | c | c | c |}
\hline
$N/M$ & $m$ & $0$ & $1$ & $2$ \\
\hline
4 & $V\left(0,0,0,m\right)$ & $\frac{17}{32M}$ & $\frac{8}{32M}$ & $-\frac{1}{64M}$ \\
\hline 
4 & $V\left(0,0,m,m\right)$ & $\frac{17}{32M}$ & $\frac{7}{32M}$ & $\frac{1}{64M}$ \\
\hline
2 & $V\left(0,0,0,m\right)$ & $\frac{1}{2M}$ & $\frac{1}{4M}$ & $0$ \\
\hline 
2 & $V\left(0,0,m,m\right)$ & $\frac{1}{2M}$ & $\frac{1}{4M}$ & $0$ \\
\hline
\end{tabular}
\caption{Approximate analytical values of $V\left(m_1,\ldots,m_4\right)$ (see Eq.~(\ref{eq:def_V})) for the dominant matrix elements in the wave packet transform of a local interaction. In the following, we us the matrix elements for $N/M=2$.}
\label{tab:V_values}
\end{table}

\section{Wilson-Wannier basis and fermion pairing}
\label{sec:wwpairing}

Symmetry breaking in fermion systems can often be understood as a transition from free to paired fermions. The best known example is superconductivity, where electrons bind into pairs which form the condensate that characterizes the superconducting state. However, spin and charge density waves may also be viewed as pairing of electrons and holes, so that a wide variety of states falls into the class of paired fermion states. A paired state introduces an energy scale $\Delta$, given by the fermion gap, and a length scale, the pair size $\xi$. In the weak coupling limit, we can estimate
\begin{equation}
\xi \approx \frac{2\pi v_F }{\Delta}
\end{equation}
on dimensional grounds.

In this section we discuss fermion pairing in the context of the WW basis. Since the WW basis states are localized on the length scale $M$, one expects that for $M >\xi$ pairs are (predominantly) local in the WW basis, whereas for $M <\xi$ they are non-local. On the other hand, the pair correlations decrease as one moves away from the Fermi surface, and the corresponding width in momentum space is $2\pi/\xi \sim \Delta/ v_F$. Hence, we expect states that with distance less than $\Delta / v_F$ to be strongly correlated, whereas they are expected to be weakly correlated when they are far away from the Fermi surface.

These estimates suggest that it is possible to replace fermionic degrees of freedom by pairs that are local (in real space) in the WW basis when $M/\xi$ is chosen large enough. In this way the low energy problem may be bosonized. Moreover, only about $\xi/M$ states in the direction perpendicular to the Fermi surface are strongly correlated, the remainder may be treated perturbatively. It is natural to expect that for $M\sim \xi$, pairs are reasonably localized in both momentum and real space, hence allowing for a simplified description of the low energy physics in terms of relatively few WW basis functions.

The remainder of this section elaborates on these heuristic considerations. In Sec.~\ref{sec:fermion_pairing}, we define fermion pairing in term of properties of dominant eigenvectors of the two-particle density matrix, exemplified by the ground state properties obtained using exact diagonalization. We also consider the consequences for the WW representation of these eigenvectors. In order to complement the analysis of small systems, we consider properties of mean-field trial wave functions and the corresponding mean-field Hamiltonian as well.

\subsection{Fermion pairing}
\label{sec:fermion_pairing}

The pairing of fermions in a given state (or density matrix) can be computed from the particle-hole (or two-particle) density matrix . For sake of concreteness, we focus on antiferromagnetic correlations, i.e. particle-hole pairing in the spin-channel. In momentum space representation, the particle-hole density matrix (PHDM) in the spin-channel is given by
\begin{equation}
\tilde{P}_q\left(p, p'\right) = 3 \left\langle \left(\tilde{S}^z_{q}\left(p\right)\right)^\dagger\, \tilde{S}^z_{q}\left(p'\right)\right\rangle,
\label{eq:def_spin_density_matrix}
\end{equation}
where the spin operators $\tilde{S}^z_q(p)$ are defined by
\begin{equation}
\tilde{S}^z_q(p) = \tilde{c}^\dagger_{p+q/2,s}\,\sigma^z_{ss'} \,\tilde{c}^\pdag_{p-q/2,s'}.
\end{equation}
The momentum $q$ is the total momentum of the operator. $p$ is the relative momentum. By virtue of translational invariance, the PHDM is diagonal in $q$. Moreover, $\tilde{P}_q(p,p')$ is hermitian, so that it can be diagonalized, yielding eigenvectors of the form $\tilde{f}_q(p)$. The Fourier transform of $\tilde{f}_q(p)$ w.r.t.~$p$ gives the shape of a particle-hole pair, and can be used to obtain the pair size $\xi$.

If the system has at least short range antiferromagnetic order, the spin-density matrix is dominated by eigenvectors that have total momenta around $\pi$. In the following we shift the total momentum by $\pi$, i.e. $q\rightarrow \pi+q$, to take this into account. The pair wave function is expected to be localized in real space on scale $\xi$. On scales larger than $\xi$ we can then speak of magnetic moment (or pair) formation, and seek to describe the low energy physics in terms of the pair degrees of freedom. 

We now turn to the form of the dominant eigenvectors in the WW basis. We consider the WW transform of the operator corresponding to an eigenvector, so that the standard formulas from Sec.~\ref{sec:wwtrafo_1} can be used. Spin indices will be omitted for sake of brevity. The WW transform of $\tilde{f}_q(p)$ is given by
\begin{equation}
\begin{split}
&\sum_{p} \tilde{f}_{\pi+q}(p) c^\dagger_{p+\pi/2+q/2} c_{p-\pi/2-q/2} =  \\ &\qquad \sum_{mk,m'k'} \frac{1}{2\sqrt{H_k H_{k'}}}\gamma^\dagger_{mk}\,\gamma^\pdag_{m'k'} \\& \qquad\times \sum_{\alpha,\alpha'} e^{-i\left(\alpha \phi_{m+k} - \alpha' \phi_{m'+k'}\right)} \bar{f}_{mk;\,m'k'},
\end{split}
\label{eq:wigner_bilinear_def}
\end{equation}
where 
\begin{equation}
\begin{split}
&\bar{f}_{mk;\,m'k'} = \sum_p \tilde{f}_{\pi+q}(p) \\  &\qquad \times \tilde{g}_{mk}^\ast(p+\pi/2 +q/2)\,\tilde{g}_{m'k'}(p-\pi/2-q/2) 
\end{split}
\label{eq:wigner_wave_packet_def}
\end{equation}
is the wave packet transform of $\tilde{f}_q(p)$. Now we assume that $Mq \ll \pi$, so that modulations of the AF order occur only on length scales that are large compared to $M$. Based on this assumption, we evaluate (\ref{eq:wigner_wave_packet_def}) to leading order in $Mq$. The result is
\begin{equation}
\begin{split}
\bar{f}_{mk;\,m'k'} \approx & e^{i M  q\left(m+m'\right)/2} \\ &\times\sum_p \tilde{f}_{\pi}(p) \,\tilde{g}_{mk}^\ast(p+\pi)\,\tilde{g}_{m'k'}(p), 
\end{split}
\end{equation}
where we have taken into account that the $q$-dependence contained in $\exp\left[iM \left(m+m'\right)q\right]$ can never be neglected since $m+m'$ can be arbitrarily large. The remaining momentum sum can be performed using the gradient expansion (\ref{eq:wave_packet_gradient_expansion}). To leading order we obtain
\begin{equation}
\begin{split}
&\sum_p \tilde{f}_{\pi}(p) \,\tilde{g}_{mk}^\ast(p+\pi)\,\tilde{g}_{m'k'}(p) \approx \\
%&\approx\\&\qquad \tilde{f}_\pi\left(K\frac{k+k'}{2}\right)  \sum_p \,\tilde{g}_{mk}^\ast(p+\pi)\,\tilde{g}_{m'k'}(p)\\
& \qquad \tilde{f}_\pi\left(K\frac{k+k'}{2}\right) \, \left(-1\right)^{m+k} \braket{g_{m,M-k}}{g_{m',k'}}.
\end{split}
\end{equation}
Inserting the result into (\ref{eq:wigner_bilinear_def}), we obtain the final expression
 \begin{equation}
 \begin{split}
& \sum_{p} \tilde{f}_{\pi+q}(p) c^\dagger_{p+\pi/2+q/2} c_{p-\pi/2-q/2} \approx \\
%\sum_{mk,m'k'} \gamma^\dagger_{mk}\,\gamma^\pdag_{m'k'} \,e^{i M  q\left(m+m'\right)/2} \tilde{f}_\pi\left(K\frac{k+k'}{2}\right) \\&\qquad \times\,\left(-1\right)^{m+k}e^{i M q(m+m')/2}\,\braket{m,k+M/2}{m,k'-M/2}\\
 &\qquad e^{i M  q\left(m+m'\right)/2} \sum_{mk} \gamma^\dagger_{m,k+M/2}\,\gamma^\pdag_{m,k-M/2} \, \\ &\qquad \times \left(-1\right)^{m+k}\tilde{f}_\pi\left(Kk\right)
 \end{split}
 \label{eq:eigenvector_ww_trafo}
 \end{equation}
 
The main conclusion to draw from this exercise is that the PHDM eigenvectors that correspond to pair formation are diagonal in the WW basis representation, provided that two conditions hold: The size $\xi$ of a pair should be less than $M$, so that the gradient expansion can be used. The range of relevant total momenta $q$ should satisfy $Mq\ll 2\pi$, i.e. only modulations of the condensate that are larger than $M$ are faithfully represented. When these conditions are met, the result (\ref{eq:eigenvector_ww_trafo}) suggests that the low energy physics of the system can be treated in a reduced Hilbert space, that contains only the pair degrees of freedom. 
% as discussed in Sec.\note{section on Hilbert space}.

\subsection{WW representation of the AF mean-field Hamiltonian}

Having discussed the general form of the PHDM for systems with (at least) short range AF order, we now illustrate the interplay of the two length scales $\xi$ and $M$. We are especially interested in the localization of the eigenvector in phase space. Since a ground state wave function is needed for the analysis, and we aim to elucidate general features only, we use the ground state of the AF mean-field Hamiltonian for this purpose, which allows to extract information for arbitrary parameters and for large systems. The Hamiltonian is given by
\begin{equation}
\begin{split}
\ham{AF} =& -t\sum_{j}\sum_s \left[ c^\dagger_{j,s} \, c^\pdag_{j+1,s} + c^\dagger_{j+1,s}\,c^\pdag_{j,s} \right] \\ & + \Delta \sum_j \left(-1\right)^j\,  c^\dagger_{j,s}\, \sigma^z_{ss'}\,c^\pdag_{j,s'},
\end{split}
\label{eq:AF_Hamiltonian}
\end{equation}
where $\Delta$ is the mean-field for the staggered magnetization, which we take to point into the $z$-direction. The Hamiltonian (\ref{eq:AF_Hamiltonian}) can be solved exactly, and the dominant eigenvector of the PHDM that describes the condensate is given by the anomalous part of the equal-time \emph{one-particle} Green's function. 

The Green's function in the ground state is given by
\begin{equation}
\begin{split}
\tilde{F}(p) &=  \sigma^z_{ss'} \left\langle \tilde{c}^\dagger_{p+\pi/2,s}\,\tilde{c}^\pdag_{p-\pi/2,s'}\right\rangle \\
&= \frac{\Delta}{\sqrt{\epsilon\left(p+\pi/2\right)^2 + \Delta^2}},
\end{split}
\label{eq:anomalous_greens_function}
\end{equation}
where $\epsilon(p) = -2t\cos p$. The WW transform $F_{mk,m'k'}$ of (\ref{eq:anomalous_greens_function}) can be evaluated in the same way as in Sec.~\ref{sec:fermion_pairing} above, but we keep terms up to $O(1/M)$ in the gradient expansion. This leads to 
\begin{equation}
\begin{split}
F_{mk,m'k'} \approx& \left(-1\right)^{m+k} \tilde{F}\left(Kk - M/2\right) \delta_{k,k'+M} \delta_{m,m'}\,\\&+\, \frac{\pi}{4M}\left(-1\right) \tilde{F}'\left(Kk-M/2\right) \\ & \times\delta_{k,k'+M} \delta_{m,m'}\left(\delta_{m,m'+1} +\delta_{m,m'-1}\right) 
\end{split}
\label{eq:anomalous_greens_function_WW}
\end{equation}

\begin{figure}
\centering
\includegraphics[height=4.5cm]{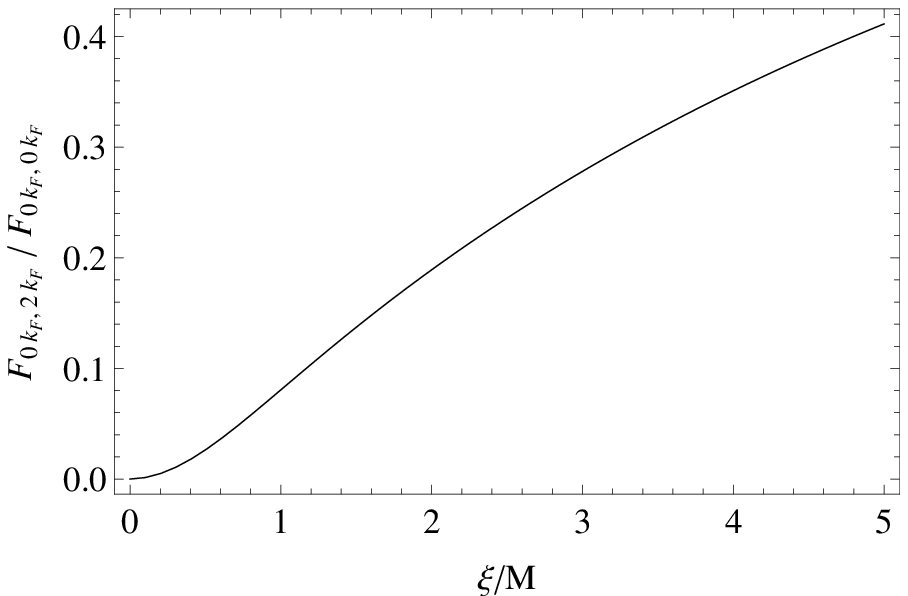}
\includegraphics[height=4.5cm]{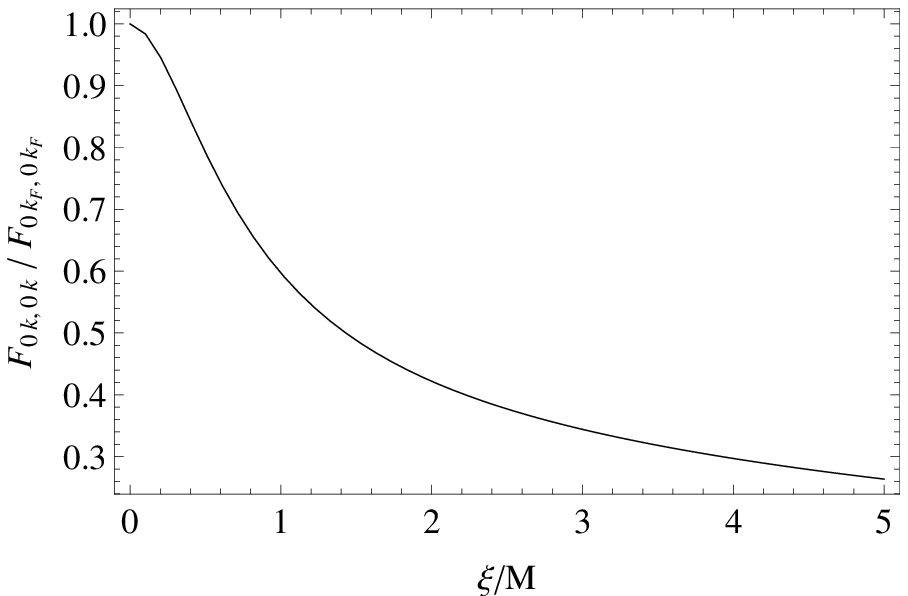}
\caption{Decay properties of the anomalous Green's function $F_{mk,m'k'}$ as function of $\xi/M$. Left panel: Real space decay measured by the ratio of $F_{0 k_F, 2 k_F}/ F_{0 k_F, 0 k_F}$, the leading non-local term divided by the local term, evaluated at the Fermi surface. Right panel: Ratio of local pair correlations at $k=k_F+1$ and $k=k_F$.}
\label{fig:sc1_k_decay}
\end{figure}

Fig.~\ref{fig:sc1_k_decay} displays the dependence of the phase space localization of the anomalous Green's function on the ratio $\xi / M$. Two quantities are of interest: First, we demand that $F_{mk,m'k'}$ should be as local in $m-m'$ as possible, so that particle-hole pairs can be considered as approximately local. Second, we want $F_{mk,m'k'}$ to decay rapidly as $k$ moves away from the Fermi surface, so that there are as few degrees of freedom as possible. Since the first criterion is improved for larger $M$, whereas the second one is optimized for small $M$, there is an optimal range $M\sim \xi$ where both are satisfied reasonably well.

The properties of the ground state Green's function (\ref{eq:anomalous_greens_function_WW}), in particular the parametric dependence of the off-diagonal matrix elements on $1/M$ suggest to split the Hamiltonian into two parts: The first part is $O(M^0)$ and contains the diagonal part of the hopping operator and the mean-field term. The second part consists of the remaining higher order hopping terms. If $M$ is chosen large enough, the former part dominates and can be used as a starting point for approximations. In particular, the $O(M^0)$ part of the Hamiltonian conserves a subspace of the full fermionic Hilbert space that contains the dominant PHDM eigenvectors discussed above in Sec.~\ref{sec:fermion_pairing}. 

The WW transform  of the mean-field term is given by
\begin{equation}
\begin{split}
&\Delta \sum_j \left(-1\right)^j\, \sigma^z_{ss'}\, c^\dagger_{j,s}\, c^\pdag_{j,s'} = \\
&\qquad \Delta \sum_{mk} \left(-1\right)^{m+k}\, \gamma^\dagger_{mk,s}\,\sigma^z_{ss'}\, \gamma^\pdag_{m,M-k,s'}.
\end{split}
\end{equation}
In general, the staggered magnetization couples the two WW orbitals $\ket{m,k}$ and $\ket{m,M-k}$. However, at the Fermi points we have $k=M/2=M-k$, so that only one orbital is involved.

Using Eq.~(\ref{eq:ham_kin_ww_1}) to transform the hopping operator, the full Hamiltonian in the limit of large $M$ is given by
\begin{equation}
\begin{split}
\ham{AF} = &\sum_{mk}\Big[ \delta_{ss'} \epsilon\left(Kk\right) \gamma_{mk,s}^\dagger\, \gamma^\pdag_{mk,s'} \\ &+\Delta\,\sigma^z_{ss'} \, \gamma^\dagger_{mk,s} \,\gamma^\pdag_{m,M-k,s'}\Big] + O\left(\frac{1}{M}\right)
\end{split}
\label{eq:ham_af_1d}
\end{equation}

In order to estimate the effect of the neglected $O(1/M)$ hopping terms, we first obtain the single particle gap $E_k$ for each pair $\ket{m,k}, \ket{m,M-k}$ of WW orbitals from the local Hamiltonian (\ref{eq:ham_af_1d}). It is given by
\begin{equation}
E_k = \sqrt{\epsilon\left(K k\right)^2 + \Delta^2}.
\label{eq:E_k_BCS}
\end{equation}
Now we compare the single particle energy with the band width $4t_k$, where the hopping rate $t_k \sim \frac{\pi}{4} v_F/M $ is given by the $k$-diagonal nearest-neighbor hopping matrix element $T(m,k;\, m+1,k)$ (cf. Sec.~\ref{sec:ww_trafo_hopping_1}). This yields the dimensionless ratio
\begin{eqnarray}
\frac{4t_k}{E_k} &\sim& \frac{\pi}{M} \frac{v_F}{\sqrt{\epsilon\left(Kk\right)^2+\Delta^2}} \nonumber \\
&\sim& \frac{\xi}{2M}\frac{1}{\sqrt{\frac{\epsilon\left(Kk\right)^2}{\Delta^2}+1}},
\end{eqnarray} 
where we have used $\xi \sim 2\pi v_F /\Delta$. It is clear that the importance of the hopping term decreases as one moves away from the Fermi points since $\epsilon\left(K k\right) \sim K v_F \left(k - p_F/K\right)$. Thus we consider the states at the Fermi points, $k \approx p_F/K$ to estimate the importance of the hopping term. When the gap $E_k$ exceeds the band width $4 t_k$, the system can be considered to be strongly coupled in the sense that the hopping term leads to corrections that can be treated perturbatively and decay over distances of about $M$. On the other hand, when $E_k < 2t_k$, the energy gain from delocalizing an electron is large enough to overcome the single particle gap locally. In this case perturbation theory around the local Hamiltonian is not expected to converge rapidly.

%%%%%%%%%%%%%%%%%%%%%%%%%%%%%%%%%%%%%%%%%%%%%%%%%%%%%%%%%%%%
%%%%% APPLICATION
%%%%%%%%%%%%%%%%%%%%%%%%%%%%%%%%%%%%%%%%%%%%%%%%%%%%%%%%%%%%

\section{Renormalization group and effective Hamiltonians}
\label{sec:wwrg}

In this section we apply the WW basis states to two strongly coupled fixed point of the RG flow for the Hubbard chains with repulsive interactions at half-filling. The low-energy phenomenology of this system is very well understood (see e.g. \cite{giamarchi, gogolin}), so that we can compare the results obtained from the wave packet approach with exact solutions that are obtained from bosonization and Bethe ansatz \cite{luther}. We do not aim at quantitative results, and merely seek to obtain qualitative features of the low-energy physics. The main concern in this respect is the reproduction of the algebraic decay of the spin correlation function. Since the WW basis breaks the translational invariance of the system, it is not obvious that power-law correlations can be obtained at all.

The qualitative nature of the study is reflected in the approximations used: Throughout, we discard all basis states except the ones at the Fermi points, with $k = p_F/K$, where $p_F$ is the Fermi momentum. We use the fixed point Hamiltonians obtained from one-loop RG for the interaction, and expand around the strong coupling limit. Despite of the simplicity of the approximation, we show that the asymptotic behavior of correlation functions is reproduced, so that the present setup can be used as the starting point for improved approximation schemes.

\subsection{Low energy parametrization of the Hamiltonian and WW representation}

The existence of the Fermi sea for interactions that are not too strong restricts the low energy degrees of freedom to narrow momentum space regions around the two Fermi points at $\pm p_F$, i.e. one may introduce a cutoff $\Lambda$ such that only states with $|p \pm p_F| < \Lambda/v_F$ are taken into account for the low energy dynamics. At weak coupling, the width of these intervals $\Lambda/v_F \ll 2p_F$. Consequently, it makes sense to split momenta as $\alpha p_F + p$, where $\alpha =\pm 1$ distinguishes left- and right-movers, and $v_F|p| \lesssim  \Lambda$ parametrizes the remaining momentum dependence. In real space, $\alpha$ describes modulation on short scales $\sim 2\pi/p_F$, whereas $p$ encapsulates modulations on length scales of at least $2\pi v_F/\Lambda \gg  2\pi/p_F$. Note that the use of $\alpha$ coincides with the one in the definition of the WW basis functions, Sec.~\ref{sec:wwbasis}. 

Mathematically, we can take the above considerations into account by parametrizing the Hamiltonian in terms of the (charge) current operators
\begin{equation}
J_{\alpha,\alpha'}\left(p',p'\right) = \sum_s c^\dagger_{\alpha p_F + p,s}\,c^\pdag_{\alpha' p_F + p'}.
\end{equation}

We treat the kinetic energy part first. For sufficiently weak interactions, we can linearize the kinetic energy around the Fermi points, so that 
\begin{equation}
\begin{split}
\ham{kin} &= \frac{2\pi v_F}{N}\sum_p \sum_\alpha \,\alpha\,p \; J_{\alpha\alpha}\left(p,p\right). 
\end{split}
\label{eq:ham_kin_momentum_space}
\end{equation}

In the spirit of the renormalization group we assume that the interaction does not depend on the momenta relative to the Fermi points, i.e. we set
\begin{equation}
\begin{split}
\ham{int} =& \frac{1}{2}\sum_{\alpha_1\cdots \alpha_4} \tilde{U}\left(\alpha_1 p_F,\ldots, \alpha_4 p_F\right) \\ & \times \delta_{\alpha_1 p_F+\alpha_2 p_F,\alpha_3 p_F+\alpha_4 p_F}  \\
&  \times \frac{1}{N}\sum_{p_1\cdots p_4}  J_{\alpha_1\alpha_2}\left(p_1,p_2\right) \, J_{\alpha_3\alpha_4}\left(p_3,p_4\right) \\ & \times \delta_{p_1+p_2,p_3+p_4},
\end{split}
\end{equation}
where $\tilde{U}\left(p_1,\ldots,p_4\right)$ is the interaction in momentum representation. Note that this approximation is analogous to the local approximation in the wave packet transformation introduced in Sec.~\ref{sec:ww_trafo_interaction_1}. Momentum conservation restricts the values of the $\alpha_i$, so that one can parametrize
\begin{equation}
\begin{split}
 \tilde{U}\left(\alpha_1 p_F,\ldots,\alpha_4 p_F\right) =& u_1\;\delta_{\alpha_1,-\alpha_3} \delta_{\alpha_2,-\alpha_4}\delta_{\alpha_1,-\alpha_2} \,   \\
& +\; u_2\;\delta_{\alpha_1,\alpha_3} \delta_{\alpha_2,\alpha_4}\delta_{\alpha_1,-\alpha_2}  \\
& +\; u_3\;\delta_{\alpha_1,-\alpha_3} \delta_{\alpha_2,-\alpha_4}\delta_{\alpha_1,\alpha_2}  \\
& +\; u_4\;\delta_{\alpha_1,-\alpha_3} \delta_{\alpha_2,-\alpha_4}\delta_{\alpha_1,\alpha_2}.
\end{split}
\label{eq:g_ology_couplings}
 \end{equation}
$u_3$ is present at half-filling only, when umklapp scattering is allowed at low energies because of $p_F = \pi/2$.

The prefactors in (\ref{eq:g_ology_couplings}) are chose such that for the case of an onsite interaction $U$ the coupling constants have the values
\begin{equation}
u_i = U.
\end{equation}

\begin{figure}
\centering
\includegraphics[width=6cm]{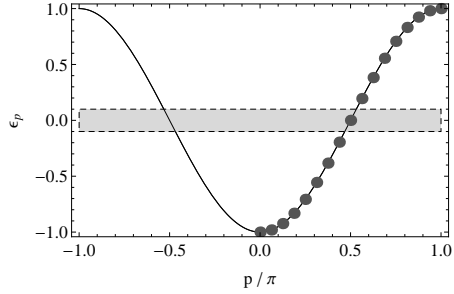}
\caption{The simplest wave packet approximation for chains: Only one WW momentum $k = p_F/K$ is kept, the remainder is discarded. It is assumed that $M$ is chosen such that only this state lies below the cutoff (shaded region). WW basis states are marked by dark dots at momentum $Kk$, and drawn on top of the dispersion of the chain.}
\label{fig:ww_chain_cutoff}
\end{figure}

Now we turn to the WW transform of the kinetic energy, (\ref{eq:ham_kin_momentum_space}), and interaction (\ref{eq:g_ology_couplings}). For sake of simplicity, we restrict the WW basis to states $\ket{mk}$ with $k=p_F/K \equiv k_F$, and assume that $M$ is chosen such that only this state lies below the cutoff, as indicated in Fig.~\ref{fig:ww_chain_cutoff}. Since $k$ is restricted to a single value $k_F$, we drop the index $k$ in the following.

The kinetic energy part in WW representation is obtained directly from (\ref{eq:ham_kin_ww_1}), leading to
\begin{equation}
\ham{kin} \approx t_{\text{eff}} \sum_s\sum_{m} \left(-1\right)^{m} \gamma^\dagger_{m,s}\,\gamma^\pdag_{m+1,s}    \; + \; \text{ h.c.},
\label{eq:ham_kin_chain}
\end{equation}
where
\begin{equation}
t_{\text{eff}}=\frac{\pi v_F}{4M}
\end{equation}

The interaction can be transformed to the WW basis using the results from Sec.~\ref{sec:ww_trafo_interaction_1}, in particular Eq.~(\ref{eq:U_ww_final_formula}), which we restate here for convenience with all indices $k_i$ dropped:
\begin{equation}
\begin{split}
U\left(m_1,\ldots,m_4\right) =& \frac{1}{2}\frac{V\left(m_1,\ldots,m_4\right)}{2} \\ &\times \sum_{\alpha_1\cdots\alpha_4} \tilde{U}\left(\alpha_1p_F,\ldots,\alpha_4p_F\right)\\
&  \times \;e^{-i\left(\alpha_1\phi_{1} +\alpha_2\phi_{2} - \alpha_3\phi_{3} - \alpha_4 \phi_{4}\right)},
\end{split}
\label{eq:U_trafo_chain}
\end{equation}
where $\phi_i = \phi_{m_i}$, and $\phi_a = \pi/2$ ($\phi_a=0$) for $a$ even (odd). The values of $V\left(m_1,\ldots,m_4\right)$ depend on the window function, the values we use are tabulated in Tab.~\ref{tab:V_values}.

Plugging the g-ology couplings (\ref{eq:g_ology_couplings}) into the right hand side of (\ref{eq:U_trafo_chain}), we observe that the Kronecker deltas can be used to perform three of the four sums over the $\alpha_i$. We evaluate the remaining sum for the $u_1$ term (the other terms being similar) only, and state the results for the other terms. We note that $\alpha_1=-\alpha_2=-\alpha_3=\alpha_4=\alpha$, and find
 \begin{equation}
\frac{1}{2} \sum_{\alpha} e^{-i\alpha\left(\phi_{1} - \phi_{2} + \phi_{3}-\phi_{4}\right)} = \cos \left[\phi_{1}+\phi_{3}-\phi_{2} - \phi_4 \right]
\end{equation}
Now recall from Eq. (\ref{eq:def_phi}) that $\phi_i$ can take on the values $0$ (for $m+k$ even) and $\pi/2$ (for $m+k$ odd) only. It follows that the cosine vanishes when an odd number of operators acts on an odd (i.e. $m+k$ odd) WW orbital, since its argument is either $\pi/2$ or $3\pi/2$. In particular, terms of the form $m_1=m_2=m_3= m_4\pm 1$ vanish, since there is always an even number of odd $k_i$ (for $k$-conserving matrix elements). Since the interactions decay rapidly with distance, the contributions from this type of interaction is thus strongly suppressed in one dimension, and we neglect them in the following. Similar cancellations occur for the other terms.

To write down the general form of the Hamiltonian for the states at the Fermi points, $k_i = p_F/K$, we define
\begin{eqnarray}
\hat{n}_m &=& \sum_s \gamma^\dagger_{m,s}\,\gamma^\pdag_{m,s}\\
\hat{s}^i_m &=& \sum_{ss'} \sigma^i_{ss'} \gamma^\dagger_{m,s}\,\gamma^\pdag_{m,s'} \\
\hat{\Delta}_m &=& \gamma^\pdag_{m\uparrow}\,\gamma^\pdag_{m\downarrow}.
\end{eqnarray}

Then the WW Hamiltonian for the states at the Fermi points is given by
\begin{equation}
\begin{split}
\ham{int}\Big|_{Kk_i=p_F} \approx& \frac{1}{M} \sum_{m} w_{\text{loc}} \, \hat{n}_m \, \hat{n}_m \\
& + \; \frac{1}{M}  \sum_{\langle m,m'\rangle} \Big[w_{\text{charge}} \hat{n}_m\,\hat{n}_{m'}\nonumber \\
&\;\;\; + \; w_{\text{spin}} \hat{\mathbf{s}}_m\cdot\hat{\mathbf{s}}_{m'}\nonumber \\
&\; \;\; + \; w_{\text{pair}} \left(\hat{\Delta}^\dagger_{m}\,\hat{\Delta}^\pdag_{m'} + \hat{\Delta}^\dagger_{m'}\,\hat{\Delta}^\pdag_{m}\right) \,\Big]
\end{split}
\label{eq:u_i_trafo_chain}
\end{equation}
with 
\begin{eqnarray}
w_{\text{loc}} &=& \frac{u_1+u_2+u_3+u_4}{8} \\ 
w_{\text{charge}} &=& \frac{-3u_1+3u_2-u_3+u_4}{32}\\
w_{\text{spin}} &=& \frac{-u_1+u_2+u_3-u_4}{32}\\
w_{\text{pair}} &=& \frac{u_1+u_2-u_3-u_4}{16}
\end{eqnarray}
where we have used the values of $V(0,0,m,m)$ from Tab.~\ref{tab:V_values}.

\subsection{Renormalization group equations}
\label{sec:rg}

We briefly review the one-loop renormalization group equations for the one-dimensional Hubbard model at weak coupling (see e.g. \cite{giamarchi}). Based on the weak coupling assumption we restrict interactions to the g-ology scheme from above.

The one-loop RG equations for the coupling constants $u_i$ are given by \cite{solyom, kopietz}:
\begin{equation}
\begin{split}
\dot{u}_1 &= -\frac{1}{\pi v_F} u_1^2 \\
\dot{u}_2 &= -\frac{1}{2\pi v_F} \left( u_1^2 - u_3^2\right)\\
\dot{u}_3 &= -\frac{1}{2\pi v_F} \left(u_1 - 2 u_2\right) u_3,\\
\dot{u}_4 &= 0,
\end{split}
\label{eq:rgflow}
\end{equation}
where the dot is shorthand for the logarithmic scale derivative $\frac{d}{ds} = -\frac{1}{\Lambda}\frac{d}{d\Lambda}$, so that $s= e^{-\Lambda/W}$. $\Lambda$ is the renormalization scale, and $W$ is the initial bandwidth. For repulsive interactions, the system of equations (\ref{eq:rgflow}) exhibits a finite scale divergence with strong coupling fixed-point given by
\begin{equation}
\begin{split}
\sqrt{2} u_2 =&  u_3 = u_{\text{AF}} > 0\\
&u_1 = u_4 = 0.
\end{split}
\label{eq:AF_fixedpoint}
\end{equation}

\subsection{WW basis analysis of RG flow}

In the RG procedure, modes above the cutoff $\Lambda$ are integrated out, yielding a renormalized Hamiltonian for the states below the cutoff. States below the cutoff are localized in momentum space around the two Fermi points in an interval of size $2\Lambda/v_F$. Since the WW basis functions are localized in momentum space, it is clear that it can be used to describe states below the cutoff. The advantage of the WW basis state over the more conventional momentum states is that the interaction are short ranged in the coarse grained real space coordinate $m$. This allows to solve the renormalized Hamiltonian exactly on small lattices, or to use strong coupling methods in order to obtain an effective Hamiltonian. On the other hand, in our experience the RG flow is much more convenient to perform in momentum space, so that we work in both bases simultaneously: We obtain the renormalized couplings in momentum space for all values of $\Lambda$ until the flow begins to diverge. From the divergence scale $\Lambda_c$ we can estimate the fermion gap, $\Delta \sim \Lambda_c$ and thus the pair size $\xi \sim 2\pi v_F/\Lambda_c$. The considerations of paired fermion states above in Sec.~\ref{sec:wwpairing} suggest that $M \gtrsim \xi$ should be chosen for the wave packet scale for a good compromise between momentum- and real-space localization. The renormalized momentum space couplings are transformed to the WW basis using (\ref{eq:u_i_trafo_chain}) for all values of $\Lambda$, and the renormalized Hamiltonian diagonalized on a small lattice (in the WW basis). Note that due to the larger lattice constant $M$ of the WW basis, the actual length scales treated are not small in general. From the exact solution as a function of $\Lambda$ we can infer the relevant low-energy degrees of freedom, and finally derive an effective model for these degrees of freedom. This model can be used to obtain the asymptotic physics at large distances. Note that the absolute scale of $M$ is unimportant in the weak coupling limit.

\begin{figure}
\centering
\includegraphics[width=6cm]{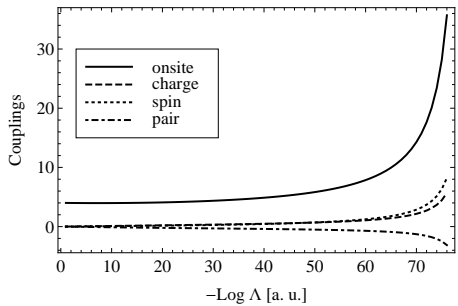}
\caption{Flow of dominant WW basis couplings for states around the Fermi points, obtained by numerical integration of (\ref{eq:rgflow}). Charge, spin, and pair refer to nearest-neighbor couplings mediated through the corresponding operators.}
\label{fig:wwcoupling_flow}
\end{figure}

Fig.~\ref{fig:wwcoupling_flow} shows the flow of the couplings constants. Instead of plotting the couplings $u_i$, the coefficients of the four terms in (\ref{eq:u_i_trafo_chain}) are shown. It is evident that the onsite repulsion is the largest coupling during the entire flow, but the nearest neighbor couplings in the charge, spin, and pair channels are non-zero as well. The flow of the coupling constants renormalizes the reduced system (\ref{eq:u_i_trafo_chain}, \ref{eq:ham_kin_chain}) consisting of the WW states at the Fermi level. 

\begin{figure}
 \centering
 \includegraphics[width=7cm]{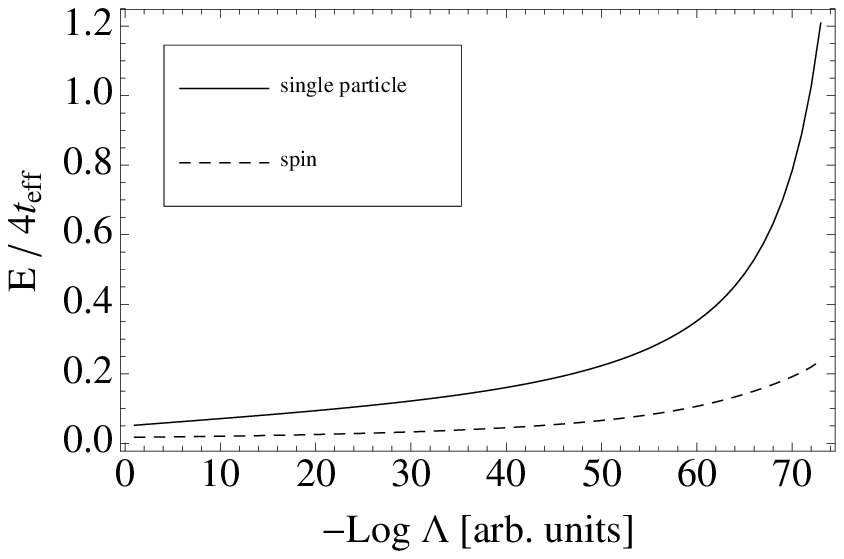}
 \includegraphics[width=7cm]{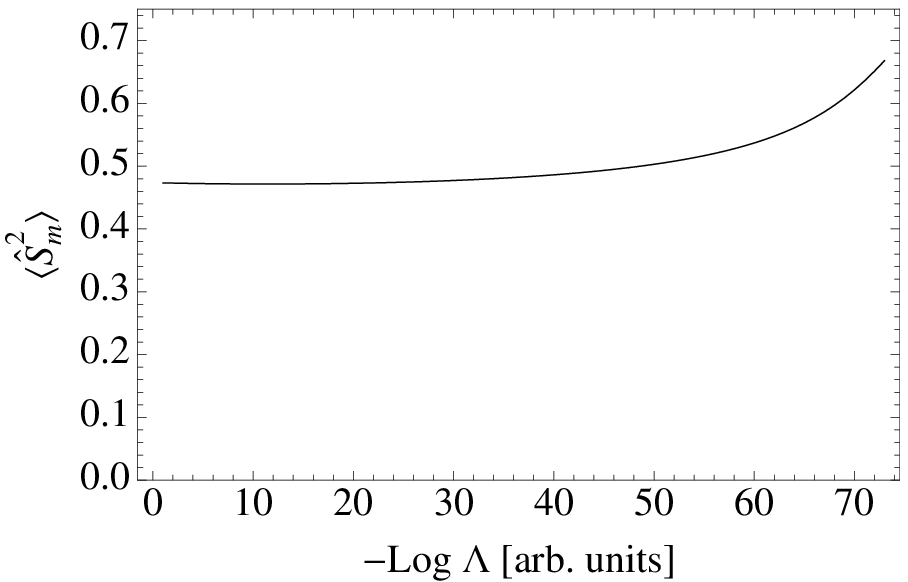}
 \caption{WW basis analysis of the renormalization group flow for half-filled Hubbard chain. Top: Flow of single particle and spin excitation energies. The flow to strong coupling is visible through the divergent single particle gap. At the same time, the spin gap remains small. Bottom: Suppression of double occupancy and emergence of local moments. At low energies, the local (in the WW basis) spin per site tends towards the maximum value $3/4$.}
 \label{fig:flowing_stuff}
 \end{figure}

Intuitively it is clear that the dominance of the onsite repulsion in the renormalized interaction Hamiltonian (\ref{eq:u_i_trafo_chain}) suppresses double occupancy, so that local moments are formed. In order to quantify this effect and to have a procedure that can handle less obvious problems as well, we propose to diagonalize the renormalized reduced Hamiltonian for a small system (here 8 sites in the WW basis with periodic boundary conditions). From the results we obtain an estimate of how strongly coupled the system is, and what the effective degrees of freedom are. The left panel of Fig.~\ref{fig:flowing_stuff} displays the effect of renormalization on the gaps for single particle and spin excitations relative to the half-filled singlet ground state. As the strong coupling regime is reached, the single particle gap exceeds the bandwidth of the reduced WW basis model, so that the weak coupling description has to be abandoned. At the same time, the spin gap stays relatively small, and evaluation of the spin per site in the ground state shows that local moments begin to form (Fig.~\ref{fig:flowing_stuff}, right panel). 
 
The above findings indicate that a good low-energy model can be based on the spin sector of the system, i.e. by projection to the two spin states $\ket{\uparrow}$ and $\ket{\downarrow}$ per WW site. Since the renormalized interaction (\ref{eq:u_i_trafo_chain}) contains a nearest-neighbor spin interaction term, the resulting model is an antiferromagnetic Heisenberg model with coupling $J = (-u_1+u_2+u_3-u_4)/32M$ when perturbative corrections from the hopping Hamiltonian are neglected, i.e.
\begin{equation}
\ham{eff} = J \sum_{m} \hat{\mathbf{s}}_{m}\cdot \hat{\mathbf{s}}_{m+1} \; + \; \text{hopping corrections}. 
\label{eq:effective_heisenberg}
\end{equation}

Since the properties of the Heisenberg model are well known, we can directly obtain the asymptotic form of spin correlation function from the effective model (\ref{eq:effective_heisenberg}). We use the Bethe ansatz solution \cite{luther,giamarchi}, which yields
\begin{equation}
\begin{split}
\left\langle \hat{\mathbf{s}}_m\cdot \hat{\mathbf{s}}_{m'}\right\rangle =& C_1 \frac{1}{\left(m-m'\right)^2}  \\&+C_2 \left(-1\right)^{m-m'} \frac{1}{|m-m'|},
\end{split}
\label{eq:spincorrelation}
\end{equation}
where the $C_i$ are constants. There are soft excitations at the points $0$ and $\pi$ of the Brillouin zone of the superlattice defined by the WW basis states. From the discussion in Sec.~\ref{sec:fermion_pairing}, especially Eq.~(\ref{eq:eigenvector_ww_trafo}) it follows that the asymptotic form (\ref{eq:spincorrelation}) in the WW basis implies the same asymptotic form in the real space lattice for momenta that are close to $0$ or $\pi$. In particular, the model reproduces the algebraic decay of the spin-spin correlation functions with the correct power laws (see e.g.~\cite{giamarchi}).

\section{Conclusions}

In this work have introduced the Wilson-Wannier basis \cite{wilsonbasis,daubechies,discretewilson}, consisting of phase-space localized basis functions, for the description of interacting fermion systems. We have reviewed the mathematics behind orthogonal basis functions, and have derived approximate transformation rules that allow for a convenient and systematically improvable application of the basis. We have also shown how these functions facilitate the derivation of low-energy models when there are strong short-ranged correlations in the form of large eigenvalues of the two-particle (or particle-hole) density matrix. 

We have focussed on one particular system, the Hubbard chain with repulsive interactions at half-filling in order to be able to introduce the relevant concepts in a concrete example. We have shown that the low-energy physics can be extracted from a combination of renormalization group methods, diagonalization of small clusters in a reduced WW basis, and the projection to an effective spin model. 

Throughout we have remained on a qualitative level in order to expose the underlying physical ideas, which led us to consider the simplest approximation wherever possible. However, since all approximations can be improved systematically, we believe that a variety of numerical and analytical techniques can be implemented to arrive at more quantitative results. In particular, the underlying Hamiltonian framework enables the use of methods such as strong coupling perturbation theory, real space renormalization group methods \cite{morningstar, dmrg}, or exact diagonalization. The application of these improvements to models on ladders and on the two-dimensional square lattice will be reported in future work.

The reasoning behind our approach originates in the renormalization group, with the aim of treating the strongly coupled renormalized Hamiltonians that frequently arise in a systematic way. The main idea was to make use of the fact that the divergence scale of the flow yields a natural length scale where the original fermionic degrees of freedom cease to offer a good description of the physics. Our approach allows to investigate the physics at this scale in an unbiased manner, and without making any assumptions about the behavior at larger scales, in contrast to semi-classical expansions around a mean-field state. Finally, the WW basis can be readily generalized to quasi-one dimensional systems with more than one band as well as to cubic lattices of arbitrary dimensionality. 

%In the second part \cite{part_two} of this series we put these features to use by applying the methodology developed above to ladder systems as well as the two-dimensional Hubbard model.

\begin{acknowledgements}
The author acknowledges fruitful discussions with M. Sigrist and T.M. Rice. This study was financially supported by the NCCR MaNEP of the Swiss national fund. 
\end{acknowledgements}

\end{document}